# Observation of Intrinsic Anderson Localization in Few-Layer $ReS_2$

Shreya Paul[1], Pritam Das[1], Devarshi Chakrabarty[1], Shamashis Sengupta[2], and Sajal Dhara[1*]

[1]Department of Physics, Indian Institute of Technology, Kharagpur 721302, India
[2]IJCLab, CNRS/IN2P3, Université Paris-Saclay, Orsay 91405, France
*sajaldhara@phy.iitkgp.ac.in

**Abstract**

Electron localization phenomena are expected to play an important role in the transport properties of two-dimensional materials. Rhenium disulfide ($ReS_2$), with its narrow conduction bandwidth, is uniquely susceptible to this effect. However, extrinsic disorder caused by fabrication methods obscures inherent localization behavior arising from reduced dimensionality and degrades transport properties. We report intrinsic Anderson localization in few-layer $ReS_2$ by eliminating extrinsic fabrication-induced disorder through all-dry van der Waals assembly and suppressing interface charge trapping through a hexagonal boron nitride (hBN) gate dielectric. Temperature-dependent transport reveals a crossover from nearest-neighbor hopping to two-dimensional (2D) Mott variable-range hopping (VRH). The non-monotonic gate-voltage dependence of activation energy provides direct access to the energy-resolved band-tail density of states of the $ReS_2$ conduction band. 2D Mott VRH yields a localization length of (3.5 ± 0.1) nm, an order of magnitude larger than disorder-dominated devices, providing a quantitative characterization of intrinsic Anderson localization in $ReS_2$.

Carrier localization governs low-temperature transport of two-dimensional (2D) semiconductors, where reduced dimensionality makes even weak disorder sufficient to localize all electronic states for non-interacting electrons[1,2]. The hallmark transport signature of Anderson localization is variable-range hopping (VRH)[3,4], reflecting phonon-assisted tunnelling between spatially localized states. At higher temperatures, nearest-neighbor hopping (NNH)[5] dominates as thermal energy becomes sufficient to activate transitions between adjacent localized sites, yielding thermally activated conductance. 2D Mott VRH has been reported in $MoS_2$,[6–11] $WS_2$,[12,13] $MoSe_2$,[14] $WSe_2$,[15] and BP[16–18] with extracted localization lengths varying widely depending on sample quality and dielectric environment. Despite extensive investigation, quantitative experimental access to intrinsic Anderson localization in 2D semiconductors has been severely limited by extrinsic disorder introduced during device fabrication, which dominates transport and obscures intrinsic localization behavior.

Rhenium disulfide ($ReS_2$) is unique among transition metal dichalcogenides for Anderson localization studies. $ReS_2$ crystallizes in a distorted 1T structure arising from Peierls distortion[19], which gives rise to pronounced in-plane anisotropy[20–28], and produces an exceptionally narrow conduction bandwidth[29,30], making the disorder-to-bandwidth ratio sufficiently large to localize carriers throughout the band and rendering $ReS_2$ particularly susceptible to carrier localization. $ReS_2$ has been incorporated as the channel material in field-effect transistors (FETs)[31–33], exhibiting n-type transport with room-temperature field-effect mobilities up to 30 $cm^2V^{-1}s^{-1}$,[33] and on/off ratios exceeding 10^6,[34]. However, in these devices, significant interface charge trapping introduces extrinsic disorder that dominates transport and obscures the intrinsic localization behavior of $ReS_2$[29,34].

Here, we report temperature-dependent transport in few-layer AA-stacked $ReS_2$ FETs fabricated by all-dry transfer on pre-patterned electrodes, with hexagonal boron nitride (hBN) as the top-gate dielectric and few-layer graphene (FLG) source, drain, and gate electrodes. The all-dry assembly eliminates fabrication-induced contamination at the channel interfaces, while the atomically flat, trap-free hBN dielectric suppresses interface charge trapping that had prevented characterization of intrinsic Anderson localization in $ReS_2$. Negligible gate hysteresis, compared to back-gated devices with $SiO_2$ dielectric fabricated from the same source crystal, confirms the effective suppression of interface disorder. Transport measurements reveal a gate-dependent crossover from thermally activated NNH to 2D Mott VRH, with both regimes confined entirely within the Anderson insulating regime across the accessible gate voltage and temperature range. Above the crossover, the activation energy exhibits a non-monotonic gate-voltage dependence, providing direct experimental access to the energy-resolved band-tail structure of the narrow $ReS_2$ conduction band. Below the crossover, 2D Mott analysis yields an intrinsic localization length in sharp contrast to prior reports from disorder-dominated devices[29,34].

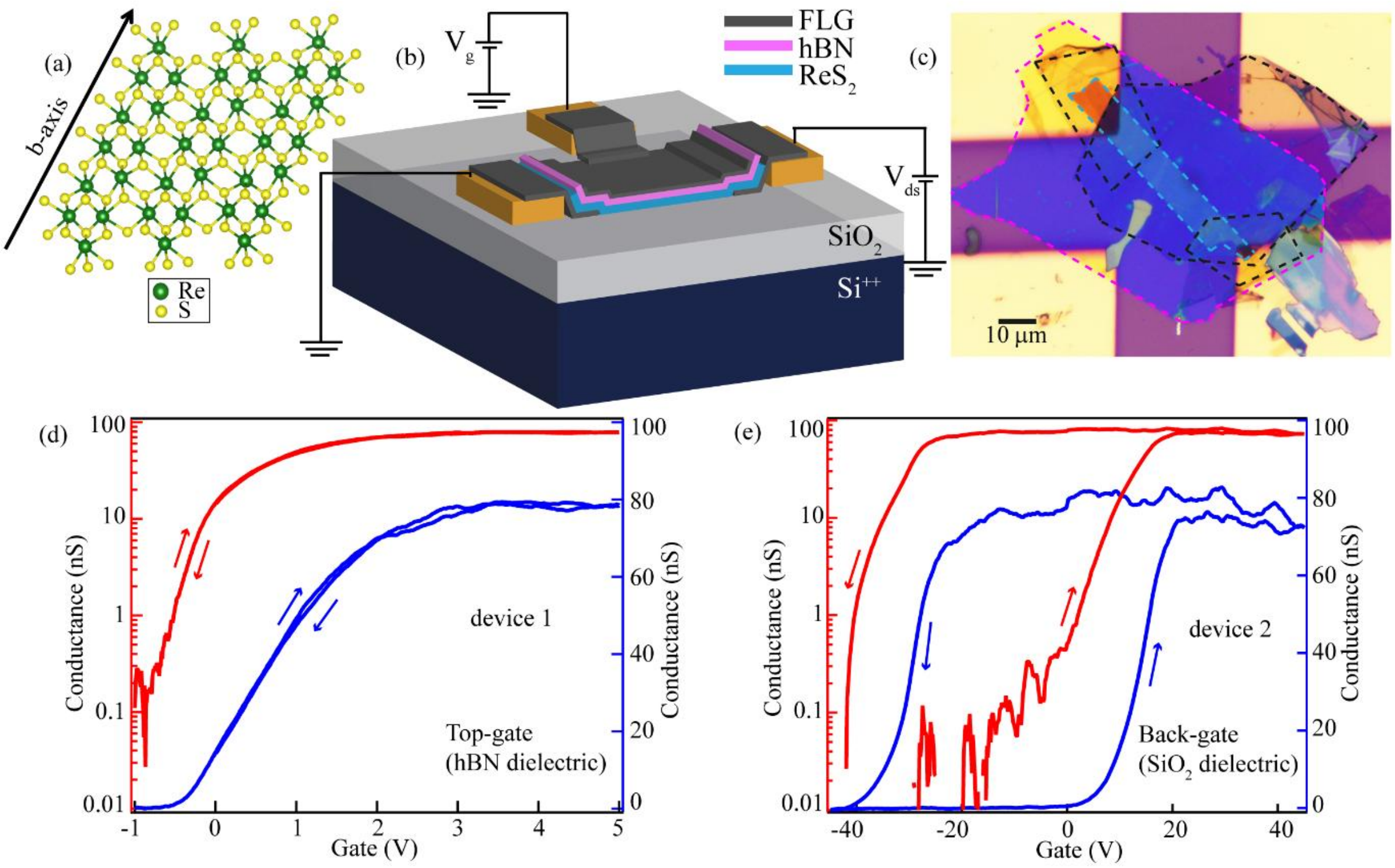

Figure 1. Van der Waals Interface Engineering in $ReS_2$ Field-Effect Transistors. (a) Top view of an AA-stacked $ReS_2$ crystal structure with the b-axis indicated by the arrow. (b) Schematic of an all-dry-transfer top-gated device architecture incorporating few-layer graphene (FLG) transparent top gate, hexagonal boron nitride (hBN) dielectric, $ReS_2$ channel, and FLG source-drain contacts on $SiO_2$/Si substrate. (c) Optical microscope image of a fabricated top-gated device showing layer boundaries of hBN (pink dashed line), $ReS_2$ layer (blue dashed line), and graphene (black dashed line). Scale bar10 µm. (d) Room-temperature transfer characteristics (conductance vs. top-gate voltage) of the top-gated FET with hBN dielectric showing negligible hysteresis ($\Delta V < 0.1V$). (e) Room-temperature transfer characteristics (conductance vs. back-gate voltage) of a separate back-gated device with $SiO_2$ dielectric, showing pronounced hysteresis ($\Delta V \approx 45V$). Source-drain bias $V_{ds}$ = 50 mV, and sweep rate 20 mV/s for both (d) and (e). Transfer characteristics are plotted on a logarithmic (left axis) and linear (right axis) scales. The arrows represent forward and reverse gate voltage sweeps.

We fabricated top-gated $ReS_2$ field-effect transistors (FETs) on prepatterned electrodes using all-dry-transfer to ensure pristine van der Waals interfaces throughout the device stack[35]. Details are provided in SI S1. A schematic of the device architecture is shown in Fig. 1(b). Mechanically exfoliated few-layer $ReS_2$ flakes serve as the channel, with Re–Re chains oriented along the crystallographic b-axis[20] (Fig. 1(a)). The $ReS_2$ channel is AA stacked, as confirmed by room-temperature Raman spectroscopy and low-temperature photoluminescence spectroscopy [SI S2]. Few-layer graphene (FLG) layers are used as source–drain electrodes which reduce the Schottky barrier height and contact resistance at the metal–semiconductor junction[36,37]. A hexagonal boron nitride (hBN) flake (~10.0 nm thickness, measured by AFM, SI S3) serves as the gate dielectric, providing an atomically flat interface with minimal charge traps and dangling bonds[38], with gate capacitance $C_{hBN}$ = 266 nF $cm^{-2}$. Transparent FLG is used as the top gate, enabling efficient electrostatic control through the thin hBN dielectric[39,40]. Figure 1(c) displays an optical microscope image of the fabricated device with distinct layer boundaries. The device geometry features channel length L ≈ 50.5 μm and width W ≈ 6.8 μm, with the channel oriented along the b-axis.

Room-temperature transfer characteristics of the top-gated device (Fig. 1(d)) demonstrate on/off ratio $\sim 10^3$ across a gate voltage ($V_g$) range of −1 to 5 V. The device exhibits n-type behavior with negligible hysteresis (ΔV < 0.1 V) between forward and reverse gate voltage sweeps, directly confirming the suppression of charge trapping at the $ReS_2$/hBN interface, consistent with the low intrinsic trap density of hBN dielectrics[41,42]. The room-temperature $I_{ds}$-$V_{ds}$ characteristics and low-temperature transfer characteristics of the top-gated device are provided in SI S4.

For comparison, room-temperature transfer characteristics of a separate back-gated device fabricated from the same source crystal using $SiO_2$ as the gate dielectric is presented in Fig. 1(e). This device exhibits pronounced hysteresis (ΔV ≈ 45 V), arising from charge trapping at the $ReS_2$/$SiO_2$ interface and within the amorphous $SiO_2$ dielectric, which contains a high density of dangling bonds, oxygen vacancies, and defect sites[42–44]. The two devices differ in layer number (~11 layers for the hBN device vs. ~23 layers for $SiO_2$ device). However , hysteresis in few-layer 2D channels is primarily governed by the gate dielectric interface rather than the channel thickness[41,42]. The order-of-magnitude contrast in hysteresis between the two devices establishes dielectric interface quality as the decisive factor controlling charge trapping in $ReS_2$ FETs, and validates the top-gated device as a clean platform for investigating intrinsic Anderson localization.

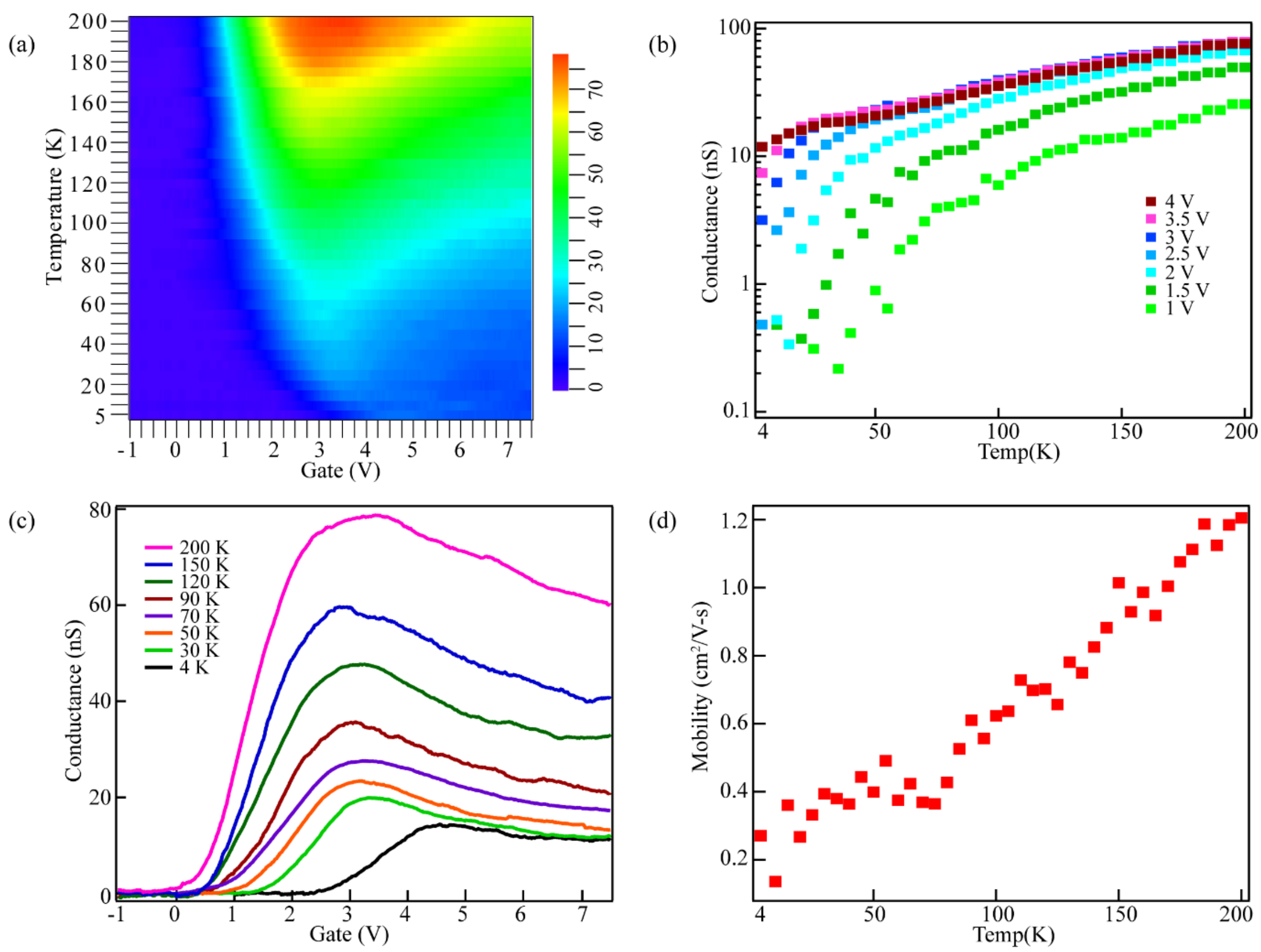

Figure 2. Gate-Dependent Crossover from Nearest-Neighbor Hopping (NNH) to Variable-Range Hopping (VRH). (a) Colour plot of two-terminal conductance versus gate voltage and temperature ($V_{ds}$ = 50 mV), showing systematic suppression with decreasing temperature. (b) Conductance (G) versus temperature for gate voltages 1- 4 V on a semi-log scale, revealing a transport regime crossover at $T^* \sim$ 100 K. (c) Conductance versus gate voltage at temperatures from 200 K to 4 K, showing a conductance maximum at $V_g \approx 2.8$ V, following suppression at higher gate voltages. (d) Field-effect mobility variation with temperature, increasing monotonically from ~0.1 cm²/V·s at 4 K to ~1.2 cm²/V·s at 200 K.

Temperature-dependent conductance (G) was measured on the top-gated device over the range of 4 to 200 K in the linear regime at a fixed bias of 50 mV. The colour plot in Fig. 2(a) shows the conductance (G) as a function of gate voltage ($V_g$) at various temperatures, revealing an overall suppression of conductance with decreasing temperature across all gate voltages. At high gate voltages ($V_g > 3$ V), the device maintains measurable conductance even at 4 K, whereas near threshold ($V_g < 1$ V), the source-drain current falls below the measurement noise floor below 50 K.

The temperature dependence of conductance on a semi-log scale (Fig. 2(b)) reveals a crossover at $T^* \approx$ 100 K. Above $T^*$, conductance decreases strongly and monotonically with decreasing temperature at all gate voltages, characteristic of thermally activated nearest-neighbor hopping (NNH). Below $T^*$, the temperature dependence weakens and becomes strongly gate-voltage-dependent, indicating a crossover to variable-range hopping (VRH) transport. This crossover at $T^*$ represents the upper bound of the NNH-to-VRH transition, and becomes apparent in Fig. 3(a).

Transfer characteristics at selected temperatures (Fig. 2(c)) show a systematic positive shift of the threshold voltage ($V_{th}$) from ≈ 0.5 V at 200 K to ≈ 2.7 V at 4 K, reflecting the progressive localization of carriers at low temperatures. The conductance reaches a maximum at $V_g \approx 2.8$ V and is partially suppressed at higher gate voltages. The partial rather than complete conductance suppression at high gate voltages is similar to the conductance dome reported in electrolyte-gated few-layer $ReS_2$[29]. In the present top-gated device, the absence of ionic gate disorder establishes this feature as intrinsic to the $ReS_2$ band structure. This suppression reflects the collapse of gate efficiency as Fermi energy ($E_F$) enters the high-DOS region of the band-tail below the mobility edge ($E_C$). This is an intrinsic feature of the narrow $ReS_2$ conduction bandwidth[29,30], discussed in the following section.

Field-effect mobility (μ) increases monotonically from ~0.1 $cm^2\ V^{-1}\ s^{-1}$ at 4 K to ~1.2 $cm^2\ V^{-1}\ s^{-1}$ at 200 K (Fig. 2(d)), yielding a positive temperature coefficient opposite to that expected for phonon-limited transport. This behavior, combined with the low mobility values, confirms that transport is governed by intrinsic carrier localization arising from the narrow $ReS_2$ conduction bandwidth, rather than band-like conduction[45]. The monotonic increase of μ with T is consistent with hopping transport observed in other localized 2D semiconductor FETs[7].

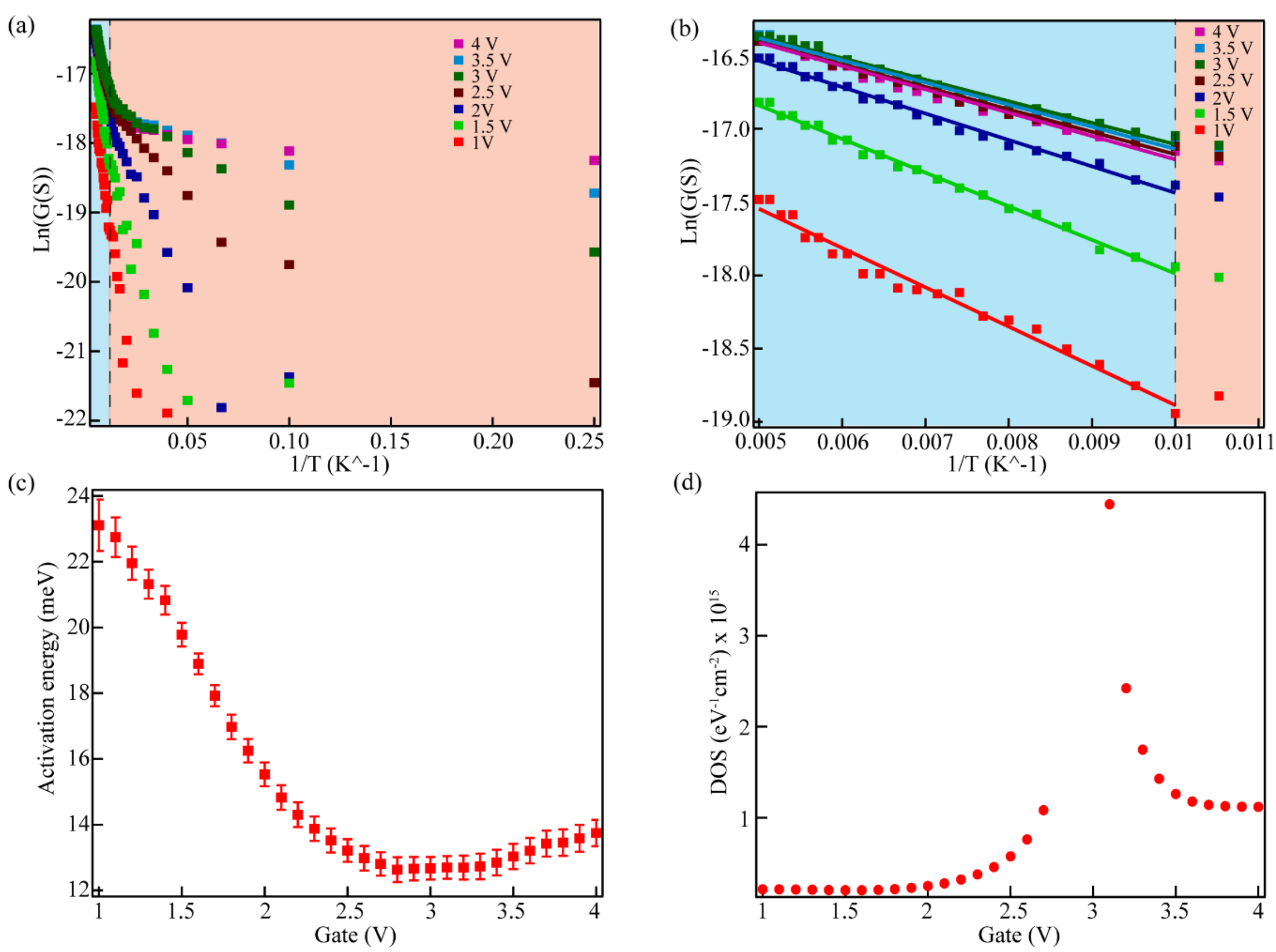

Figure 3. Band-Tail Density of States from Non-Monotonic Activation Energy. (a) Arrhenius plots (ln(G) vs. 1/T) for $V_g$ = 1– 4 V. Blue and orange backgrounds denote the high-temperature (T > 100 K) and low-temperature (T < 100 K) regimes of distinct temperature dependence. (b) Arrhenius plots zoomed over the NNH regime (T > 100 K) with linear fits (solid lines) for $V_g$= 1– 4 V. (c) Activation energy $E_a$, extracted from Arrhenius fits, as a function of gate voltage. (d) Variation of density of states $g(E_F)$ with gate voltage, extracted from the activation energy.

Figure 3(a) shows the Arrhenius plot ln($G$) versus $1/T$ for gate voltages from 1 V to 4 V, revealing two distinct transport regimes separated at $T^* \approx 100$ K. Above $T^*$(blue region), all curves exhibit prominent linear behavior characteristic of thermally activated NNH transport,

$$G(T) = G_0\, e^{-E_a/k_B T} \qquad (1)$$

where $E_a$ is the activation energy, equal to the energetic distance between $E_F$ and $E_C$. Linear fits (solid lines) in the NNH regime (Fig. 3(b)) yield $R^2 \geq 0.98$ across all gate voltages, confirming $T^* \approx 100$ K as the crossover temperature. Below $T^*$(orange region), the data deviate systematically from Arrhenius linearity, signalling a crossover to VRH transport. The activation energies extracted from Fig. 3(b) are plotted versus gate voltage in Fig. 3(c). For Vg < 2.8 V, $E_F$ moves through the band-tail states toward $E_C$, reducing $E_a$ monotonically by ~10 meV. The finite minimum at Vg ≈ 2.8 V establishes that $E_F$ reaches its closest approach to $E_C$ without crossing it. Above Vg = 2.8 V, $E_F$ enters the high-DOS region below $E_C$, where the large number of states per unit energy resists further $E_F$ movement, causing gate efficiency to collapse and $E_a$ to rise ~1 meV toward Vg = 4 V, consistent with the conductance suppression observed in Fig. 2(c). This non-monotonic gate-voltage dependence reflects the transport physics of the narrow $ReS_2$ conduction bandwidth (D ~0.4 eV)[29], which makes the disorder strength *W*/*D* sufficiently large to localize carriers throughout the band. This behavior is absent in wider-band TMDs such as $MoS_2$ and $WS_2$, where $E_F$ crosses the mobility edge within experimentally accessible gate voltage ranges, producing a metal-insulator transition[6,12,45].

In the present top-gated device gate-induced disorder is ruled out as a contributing factor, unambiguously establishing the non-monotonic $E_a$ as an intrinsic consequence of the narrow $ReS_2$ conduction band. Throughout the entire gate range, the Ioffe–Regel parameter $k_F * l \ll 1$ (SI S5), confirming that the system remains in the Anderson insulating regime.

Figure 3(d) presents the density of states $g(E_F)$ at the Fermi level[12], extracted from the gate-voltage dependence of $E_a$ using,

$$g(E_F) = \frac{C_{hBN}}{e} \left| \frac{dE_a}{dV_g} \right|^{-1} \qquad (2)$$

The quantum capacitance correction to $g(E_F)$ extracted via Eq. (2) is less than 1%, well within experimental uncertainty (SI S8). Reliable extraction is restricted to 1.2 V < $V_g$< 1.8 V, where $E_a$varies smoothly and monotonically; outside this window, particularly near $V_g \approx 2.8$ V where $\frac{dE_a}{dV_g} \rightarrow 0$, the extracted $g(E_F)$ diverges as a mathematical artifact of the vanishing derivative rather than reflecting a physical feature of the DOS. Within this window, $g(E_F)$ plateaus at $(2.06 \pm 0.04) \times 10^{14}$ $eV^{-1}$ $cm^{-2}$, in agreement with the theoretical 2D DOS $g(E_F) = g_s g_v m^*/2\pi\hbar^2 = 2.08 \times 10^{14}$ $eV^{-1}$ $cm^{-2}$ using electron effective mass $m^* = 0.5m_0$[29], spin degeneracy $g_s = 2$, valley degeneracy $g_v = 1$.

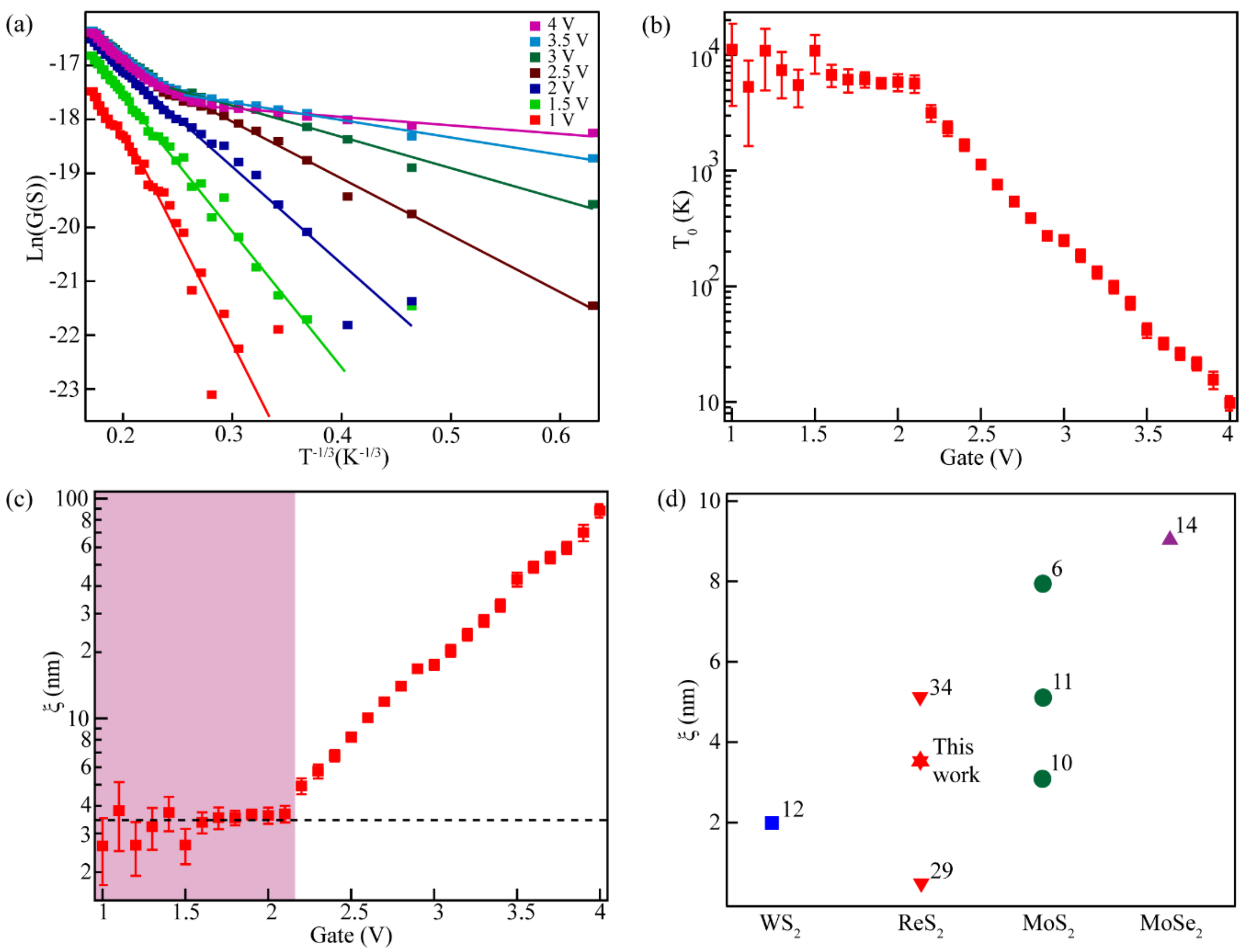

Figure 4. Intrinsic Anderson Localization Length from Two-Dimensional Mott Variable-Range Hopping (VRH). (a) VRH plots (ln(G) vs. $T^{-1/3}$) for $V_g$ = 1– 4 V with linear fits confirming 2D Mott VRH. (b) Characteristic temperature $T_0$ versus gate voltage, decreasing monotonically with increasing gate voltage. (c) Localization length ξ versus gate voltage. The dotted line indicates the intrinsic localization length ξ = (3.5 ± 0.1) nm, extracted from the shaded region. (d) Comparison of localization lengths across TMD semiconductors, with prior $ReS_2$ reports. Numbers adjacent to data points denote references.

Below $T^* \approx 100$ K, transport is governed by 2D Mott VRH,

$$G(T) = G_0(T)\, e^{-(\frac{T_0}{T})^{1/3}} \qquad (3)$$

where $T_0$ is the characteristic hopping temperature. Figure 4(a) presents linearized VRH plots (ln($G$) vs. $T^{-1/3}$) for gate voltages from 1 V to 4 V. At low carrier densities ($V_g$ < 2.5 V), VRH is well-established over the full range 10–100 K, while at high carrier densities ($V_g$ = 4 V), the clean VRH regime is confined to 10–50 K, with a mixed crossover region between ~50 K and $T^*$. The curves exhibit higher $R^2$ across all gate voltages [SI S6], confirming 2D Mott VRH as the operative transport mechanism, consistent with the finite DOS at $E_F$ established earlier.

$T_0$ decreases monotonically from ~1 ×10⁴ K at $V_g$ = 1 V to ~10 K at $V_g$ = 4 V (Fig. 4(b)), consistent with the progressive movement of $E_F$ through the band-tail states toward $E_C$ and the systematic weakening of carrier localization with increasing gate voltage. The simultaneous breakdown of both the NNH Arrhenius analysis and the VRH fits above $V_g$ = 4 V confirms that $E_F$ has entered the high-DOS region immediately below $E_C$, consistent with the band-tail structure established in the previous section.

The localization length is calculated as, $\xi = \sqrt{\frac{13.8}{k_B T_0 g(E_F)}}$ (4)

using the theoretical 2D DOS $g(E_F)$= 2.08 × $10^{14}$ $eV^{-1}$ $cm^{-2}$ uniformly across all gate voltages. ξ increases systematically from ~3nm at Vg = 1 V to ~24 nm at Vg = 3.2 V, with further increase at higher gate voltages representing upper bounds due to VRH self-consistency violation (SI S7), consistent with $E_F$ progressing toward $E_C$(Fig. 4(c)). The intrinsic localization length, extracted as the weighted mean over Vg = 1–2.1 V, is (3.5 ± 0.1) nm, satisfying $\xi \gg a$ = 0.28 nm[46] (Re–Re bond length).

Figure 4(d) compares the localization length ξ across TMD semiconductors[6,11,12]. The intrinsic localization length (3.5 ± 0.1) nm stands in sharp contrast to ξ ~ 0.35 nm[29] from disorder-dominated devices, which is comparable to the Re–Re bond length (0.28 nm[46]), directly demonstrating that van der Waals interface engineering is essential for accessing the intrinsic Anderson localization in $ReS_2$. A comprehensive comparison of localization lengths extracted from 2D Mott VRH across 2D semiconductors is provided in SI S9.

In summary, we have demonstrated intrinsic Anderson localization in few-layer $ReS_2$ field-effect transistors through van der Waals interface engineering with hBN gate dielectric, inaccessible in devices where fabrication-induced contamination and interface charge trapping dominate transport. Temperature-dependent transport reveals a gate-dependent crossover from thermally activated NNH to 2D Mott VRH, with both regimes confined entirely within the Anderson insulating regime across all accessible gate voltages. The non-monotonic gate-voltage dependence of the activation energy provides direct experimental access to the energy-resolved band-tail density of states of the $ReS_2$ conduction band. 2D Mott VRH analysis yields a localization length of (3.5 ± 0.1) nm, an order of magnitude larger than values from disorder-contaminated devices in prior reports. These results demonstrate that van der Waals interface engineering is essential for uncovering intrinsic quantum transport in narrow-band 2D semiconductors.

**Supporting Information**

Device fabrication (S1), AA-stacking identification by Raman and photoluminescence spectroscopy (S2), AFM thickness determination and gate capacitance calculation (S3), current–voltage characteristics (S4), Ioffe–Regel parameter calculation (S5), comparison of variable-range hopping models (S6), VRH parameter extraction and self-consistency analysis (S7), quantum capacitance correction to the density of states (S8), and localization length comparison across two-dimensional semiconductors (S9). Figures S1–S6, Tables S1–S3 (PDF).

**Acknowledgment**

We acknowledge the Micro Raman Spectrometer Laboratory, Central Research Facility, Indian Institute of Technology Kharagpur, for room temperature Raman spectroscopy measurements and the AFM Facility, Materials Science Centre, Indian Institute of Technology Kharagpur, for AFM measurements.

This work has been supported by funding from the Ministry of Education (MoE/STARS-1/647) and the Indian Institute of Technology Kharagpur.

**Notes**

The authors declare no competing financial interest.

**Data Availability**

The data are available from the corresponding author upon reasonable request.

Supplemental Material for

# Observation of Intrinsic Anderson Localization in Few-Layer $ReS_2$

Shreya Paul[1], Pritam Das[1], Devarshi Chakrabarty[1], Shamashis Sengupta[2], and Sajal Dhara[1,*]

[1]Department of Physics, Indian Institute of Technology Kharagpur, Kharagpur 721302, India
[2]IJCLab, CNRS/IN2P3, Université Paris-Saclay, Orsay 91405, France
*sajaldhara@phy.iitkgp.ac.in

**List of Contents:**

## S1. Device Fabrication

Few-layer $ReS_2$, hexagonal boron nitride (hBN), and few-layer graphene (FLG) flakes were obtained by mechanical exfoliation from high-quality bulk crystals (HQ Graphene) using the standard Scotch tape method and deposited on transparent PDMS stamps. Freshly cleaved flakes were inspected by optical microscopy to identify uniform, thin regions.

Top-gated van der Waals heterostructure FETs were assembled by sequential all-dry transfer [1] onto pre-cleaned Si/$SiO_2$ substrates (285 nm oxide). The stack was assembled in the following order: (i) FLG source and drain electrodes were first transferred onto the substrate on top of prepatterned metal contact pads (Ti/Au, 5/25 nm), (ii) the AA-stacked $ReS_2$ channel flake was transferred to bridge the FLG contacts, (iii) the hBN dielectric flake (~10.0 nm, confirmed by AFM, Section S3) was transferred onto the $ReS_2$ channel, (iv) a FLG top-gate electrode was transferred onto the hBN dielectric, without solvent exposure or resist contact with the channel. The crystallographic b-axis orientation of the $ReS_2$ channel was confirmed from the straight cleave edges visible in the optical micrograph (Fig. 1(c)), consistent with the preferential cleavage of $ReS_2$ along the b-axis direction.

For comparison, back-gated FETs were fabricated from the same $ReS_2$ source crystal using $SiO_2$ as the gate dielectric. The $ReS_2$ flake was transferred to bridge the metal contacts. The heavily doped Si substrate serves as the back gate.

## S2. Identification of AA stacked $ReS_2$

### S2.1 Raman Spectroscopy

Room temperature Raman measurements were performed using a 532 nm excitation laser focused through a 100x objective lens. The scattered light was collected and dispersed using a single monochromator (Horiba Scientific, 300 grooves/mm grating). The frequency separation between Mode I and Mode III is 12.8 cm$^{-1}$, as shown in Fig. S1(a), which is consistent with ~12 cm$^{-1}$ for AA-stacked $ReS_2$ [2,3].

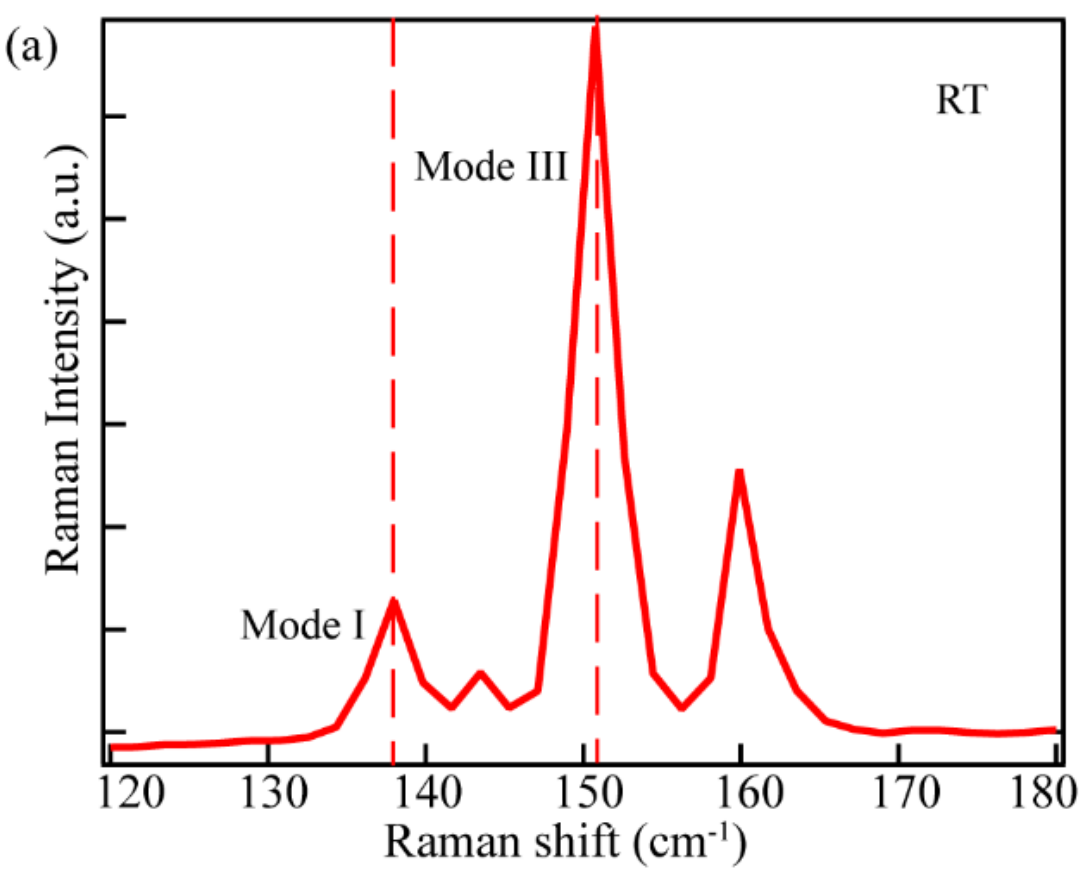

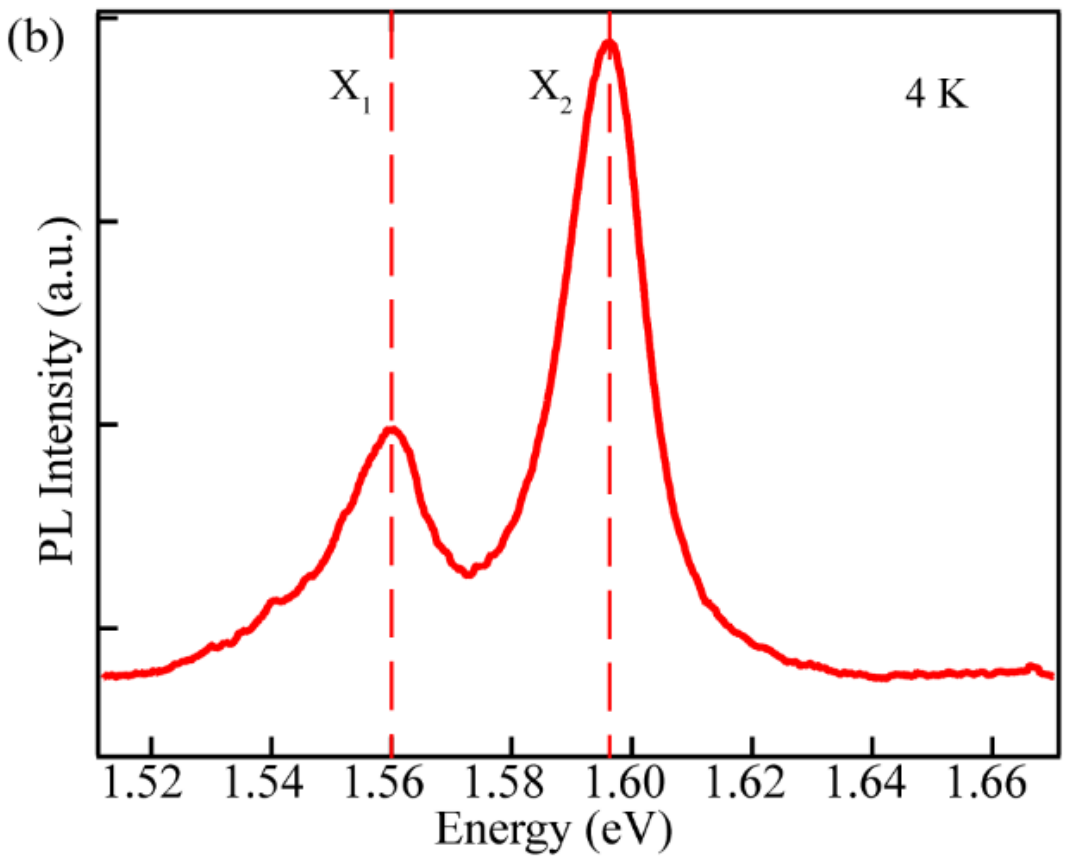

FIG S1. (a) Room temperature Raman spectra showing Mode I-Mode III separation 12.8 cm⁻¹ confirming AA-stacked $ReS_2$. (b) Low-temperature PL spectra confirming AA-stacked $ReS_2$.

### S2.2 Photoluminescence spectroscopy

Low-temperature photoluminescence (PL) measurements were performed at 4 K in a closed-cycle cryostat using a home-built confocal microscope setup. A 660 nm laser was used for excitation (spot size of ~1 μm), and the spectrometer's estimated resolution was 0.86 nm, resulting in an error bar of ~1.72 meV. Figure S1(b) shows two peaks corresponding to $X_1$ and $X_2$ excitons at (1.560 ± 0.002) eV and (1.596 ± 0.002) eV. These peaks are consistent with literature values reported for AA-stacked few-layer $ReS_2$ [4,5].

## S3. AFM-Based Thickness Determination

Atomic force microscopy (AFM) height profiles were acquired in tapping mode under ambient conditions to determine the thickness of each flake in the device stack. Line profiles extracted across the flake edges of the top-gated device and the back-gated device in main text are shown in Figure S2.

The $ReS_2$ channel flake in Fig. 1(e) in main text shows a thickness of (16.0 ± 0.2) nm (Fig. S2(a)), corresponding to approximately 23 layers (monolayer thickness ~0.7 nm [6]). The $ReS_2$ channel flake in Fig. 1(c,d) in main text shows a thickness of (8.0 ± 0.2) nm (Fig. S2(b)), corresponding to approximately 11 layers. The hBN flake Fig. 1(c) in main text shows a thickness of (10.0 ± 0.2) nm (Fig. S2(c)).

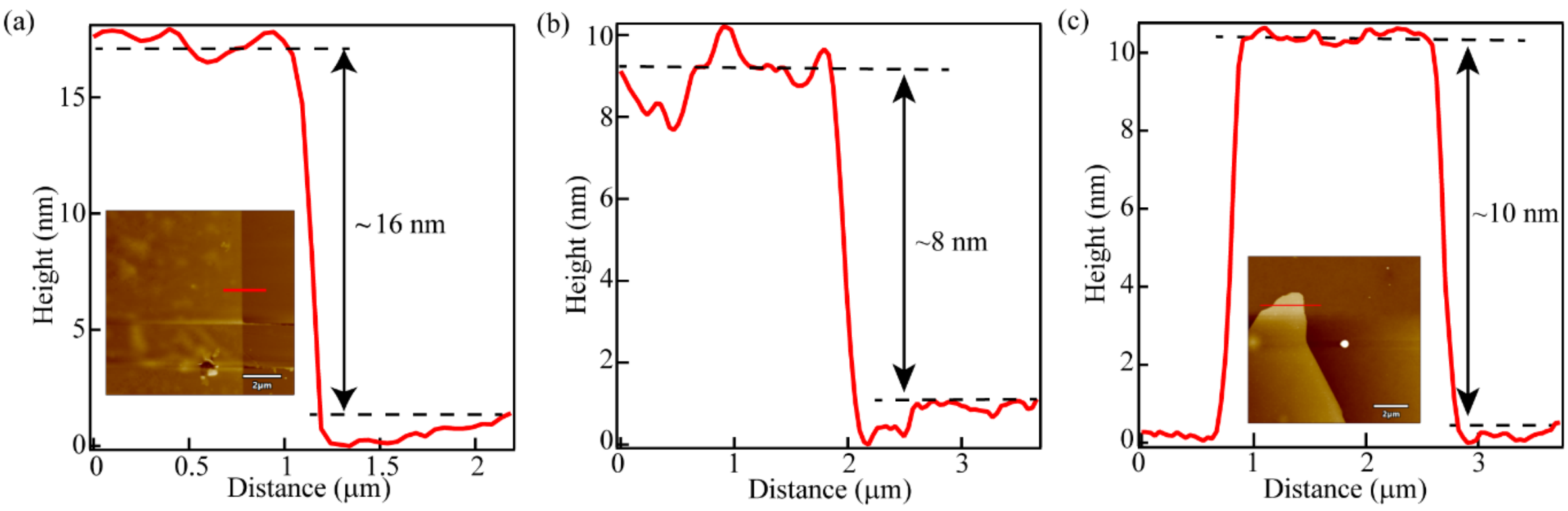

FIG S2. AFM height profiles of the flakes used in the main text. (a) Thickness of the $ReS_2$ flake in Fig. 1(e). AFM image of the flake (inset, scale bar 2 μm). (b)Thickness of the $ReS_2$ flake in Fig. 1(c,d). (c) Thickness of the hBN flake in Fig. 1(c). AFM image of the flake (inset, scale bar 2 μm).

**S3.1 Gate Capacitance Determination**

The gate capacitance per unit area of the hBN dielectric is calculated using the parallel plate formula,

$C_{hBN} = \frac{\varepsilon_0 \varepsilon_{hBN}}{d_{hBN}}$ = 266 nF $cm^{-2}$ where $\varepsilon_{hBN}$ = 3 is the out-of-plane dielectric constant of hBN [7,8] and $d_{hBN}$ = (10.0 ± 0.2) nm is the hBN flake thickness, determined by AFM.

**S4. Current–Voltage Characteristics**

**S4.1. Electrical measurements at room temperature**

All electrical measurements were performed in a closed-cycle cryostat. Temperature-dependent transport measurements were carried out from 4 to 200 K in steps of 5 K, with the sample allowed to stabilize at each temperature prior to measurement. All transport measurements are performed in the linear (ohmic) regime, where conductance $G = \frac{I_{ds}}{V_{ds}}$ is independent of the applied bias. Room temperature transfer characteristics are performed for a fixed source-drain bias $V_{ds}$ = 50 mV (in linear regime). Figure S3 represents room-temperature $I_{ds}$-$V_{ds}$ characteristics for the devices mentioned in Fig 1(d,e).

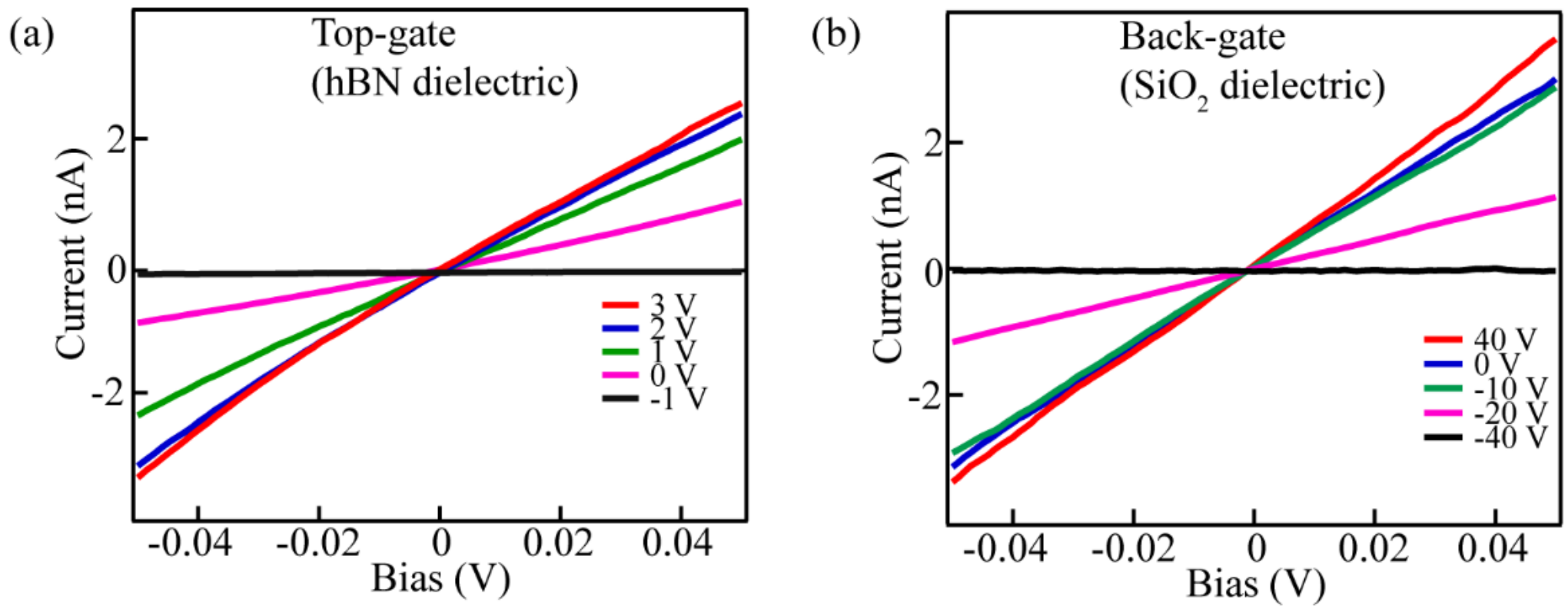

FIG S3. Room temperature I-V characteristics of $ReS_2$ FETs with (a) hBN dielectric and (b) $SiO_2$ dielectric for different gate voltages, confirming ohmic transport at room temperature. Sweep rate 20 mV/s.

## S4.2 Low-temperature transfer characteristics

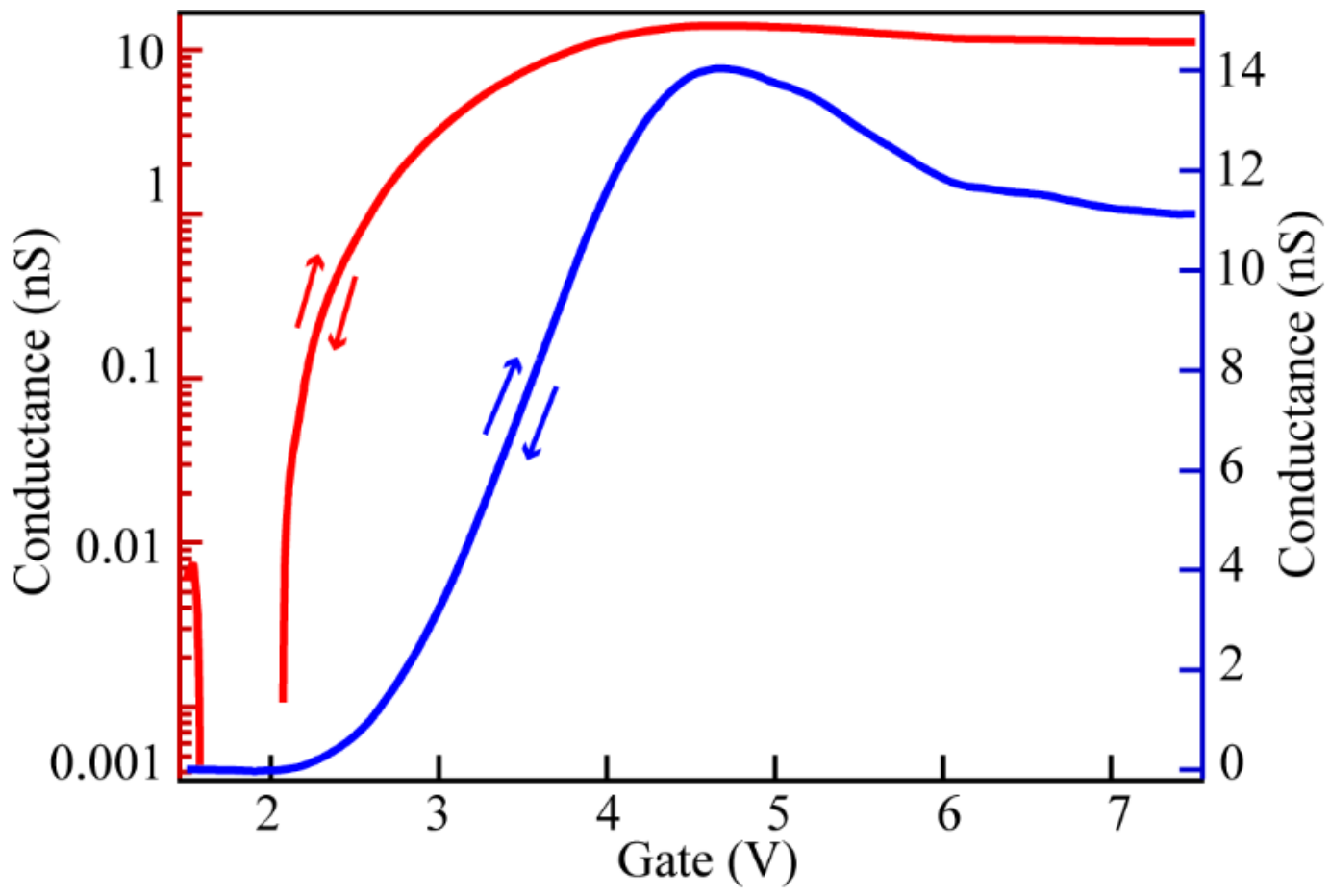

FIG S4. Transfer characteristics of the top-gated FET with hBN dielectric showing negligible hysteresis at 4 K, plotted on a logarithmic scale (left axis) and linear scale (right axis). Source-drain bias $V_{ds}$ = 50 mV, and sweep rate 20 mV/s. The arrows represent forward and reverse gate voltage sweeps.

Figure S4 shows transfer characteristics of the hBN-gated device recorded at 4 K with forward and reverse gate sweeps at 20 mV/s. The hysteresis is negligible ($\Delta V < 0.1$ V) at 4 K confirming that gate-induced disorder does not affect the intrinsic localization. Field-effect mobility was extracted as $\mu = \frac{L}{WC_{hBN}}(\frac{\partial G}{\partial V_g})$, where $L$ and $W$ are the channel length and width respectively.

**S5. Ioffe–Regel Parameter Calculation**

The Ioffe–Regel parameter $k_F * l$ defines the transport regime, with $k_F * l \ll 1$ indicating the strongly localized (Anderson insulating) regime where quantum interference dominates over semiclassical transport. Here, we calculate $k_F * l$ as a function of temperature to confirm that the present device remains in the Anderson insulating regime throughout the entire measurement. Ioffe–Regel parameter as a function of temperature is shown in Fig. S5.

The Fermi wavevector in two dimensions is related to the 2D carrier density n by:

$$k_F = \sqrt{2\pi n} \quad \text{(S5.1)}$$

At 4 K, the intrinsic carrier density $n_0$ is negligibly small compared to the gate-induced carrier density. The total carrier density reduces to:

$$n = \frac{C_{hBN}(V_g - V_{th})}{e} \quad \text{(S5.2)}$$

with $C_{hBN}$ = 266 nF cm$^{-2}$, assuming the model of geometric capacitance with metallic electrodes.

The mean free path l is:

$$l = \frac{\hbar \mu k_F}{e} \quad \text{(S5.3)}$$

The Ioffe–Regel parameter, $k_F * l = \frac{h\mu n}{e}$ (S5.4)

As a representative estimate, at $V_g$ = 3 V and T = 100 K ($V_{th}$ = 0.8 V, n = 3.66 × $10^{12}$ $cm^{-2}$, μ ≈ 0.62 $cm^2$ $V^{-1}$ $s^{-1}$), the mean free path $l$ ≈ 0.02 nm, more than one order of magnitude smaller than the Re–Re bond length (0.28 nm), confirming the system is in the strongly localized regime.

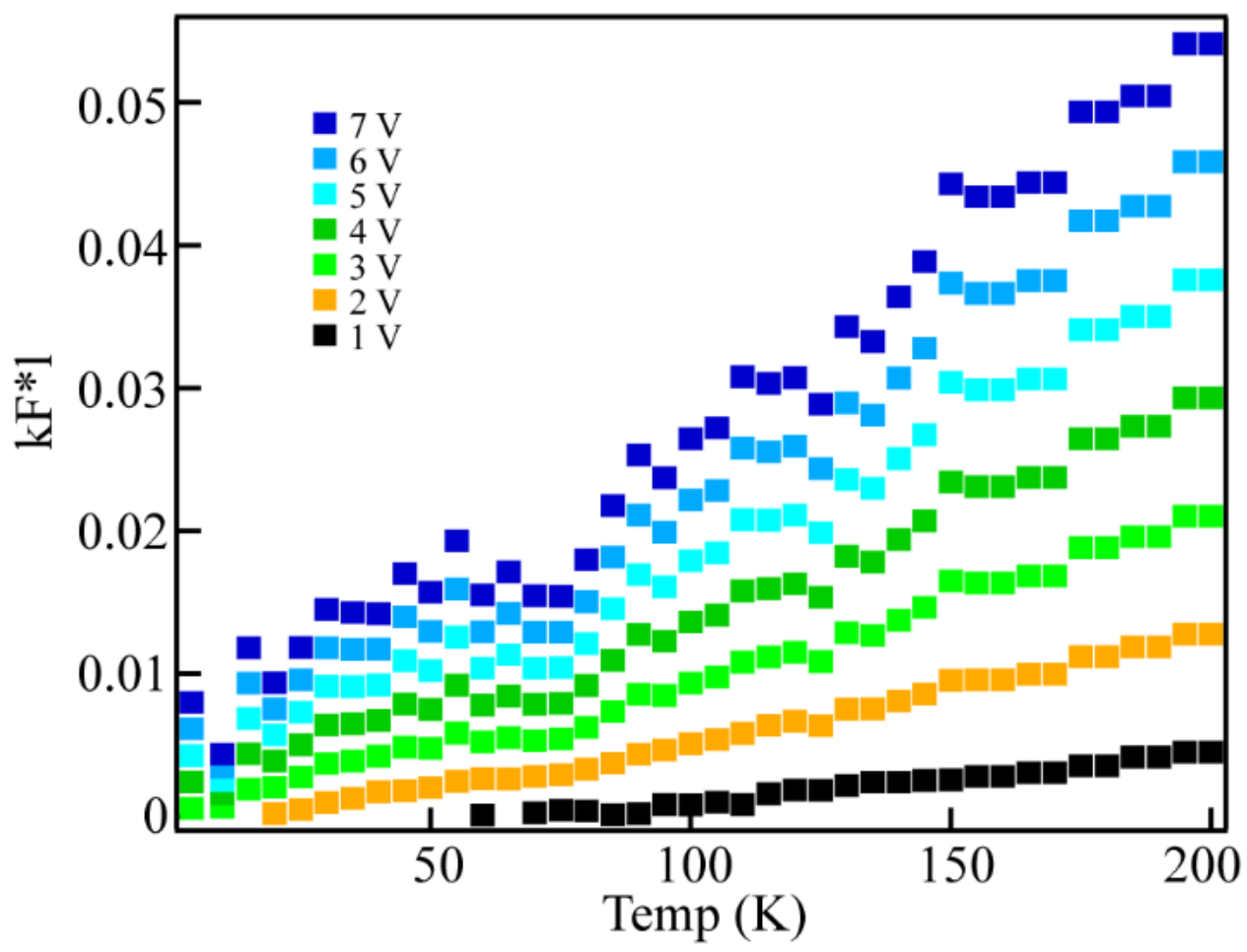

FIG S5. Ioffe–Regel parameter as a function of temperature.

**S6. Comparison of Variable-Range Hopping Models**

VRH models were evaluated against the low-temperature conductance data (T < $T^*(V_g)$) to identify the operative transport mechanism. Two models were tested, 2D Mott VRH [9] and Efros–Shklovskii (ES) VRH [10].

2D Mott VRH (hopping with constant DOS at $E_F$):

$\ln(G(T)) \propto -(\frac{T_0}{T})^{1/3}$ (S6.1)

ES VRH (hopping with Coulomb gap at $E_F$):

$\ln(G(T)) \propto -(\frac{T_{ES}}{T})^{1/2}$ (S6.2) where, $T_{ES}$ is the Efros-Shklovskii characteristic temperature.

Linear fits were performed for both models at each gate voltage. The $R^2$ values are summarized in Table S1.

Table S1. $R^2$ values for 2D Mott VRH and ES-VRH fits at each gate voltage.

| $V_g$ (V) | $R^2$ (ES-VRH) | $R^2$ (2D Mott VRH) |
|---|---|---|
| 1.0 | 0.484 | 0.541 |
| 1.5 | 0.778 | 0.823 |
| 2.0 | 0.964 | 0.975 |
| 2.5 | 0.986 | 0.994 |

| 3.0 | 0.963 | 0.983 |
|---|---|---|
| 3.5 | 0.905 | 0.941 |
| 4.0 | 0.712 | 0.779 |

2D Mott VRH yields systematically higher $R^2$ than ES-VRH at all gate voltages, confirming it as the dominant hopping transport mechanism across the full gate range.

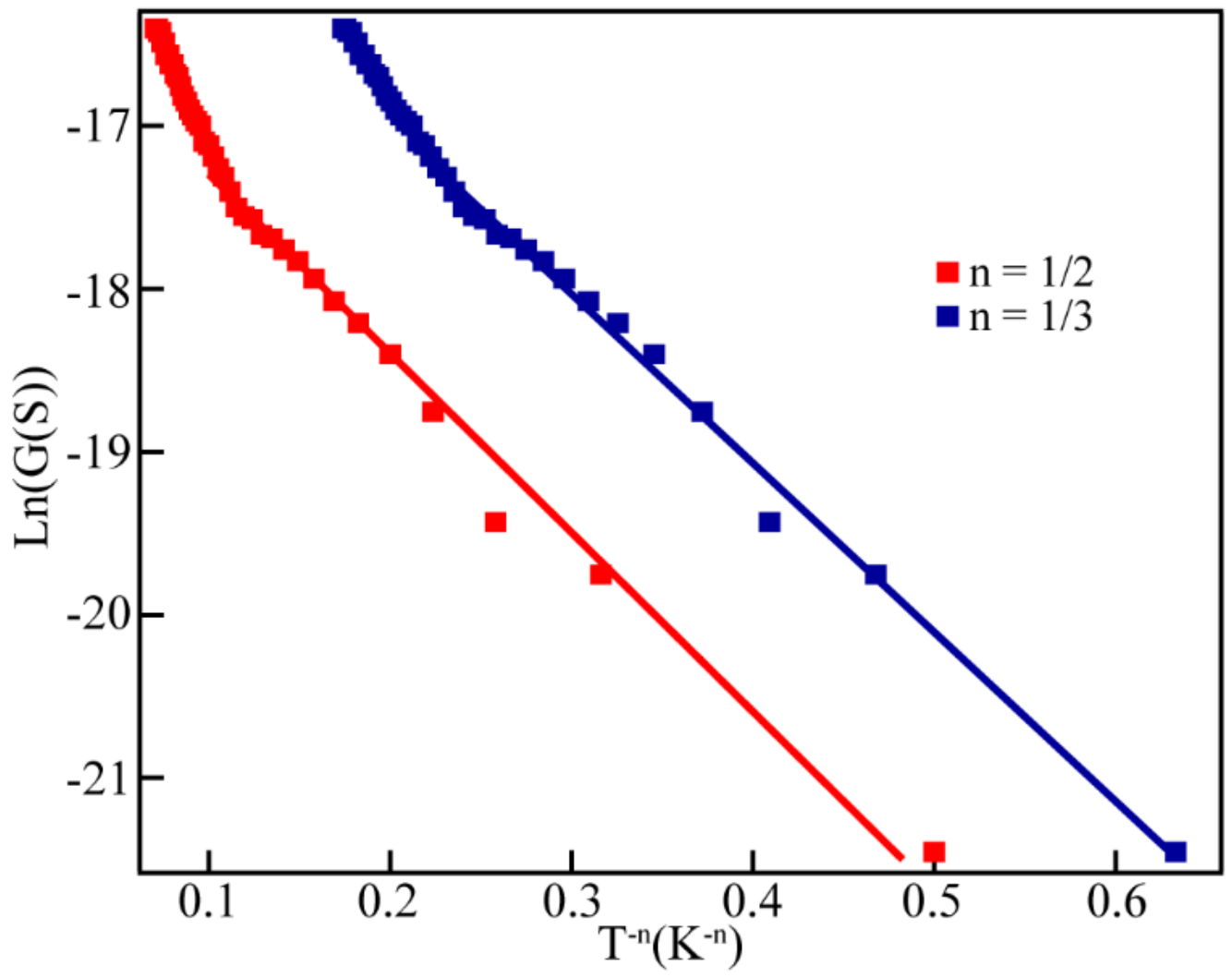

FIG S6. Comparison of 2D ES VRH (n = 1/2, red) and 2D Mott VRH (n = 1/3, blue) fits to low-temperature conductance data at $V_g$ = 4 V, demonstrating superior linearity for 2D Mott VRH over the entire temperature range 4-100 K.

Figure S6 illustrates the comparison at $V_g$ = 4 V, where the deviation of ES-VRH from linearity is most pronounced.

## S7. VRH Parameter Extraction and Self-Consistency

### S7.1 Gate-Voltage-Dependent Reliability of $T_0$ Extraction

The VRH fitting window varies with gate voltage. At all gate voltages, the upper bound is set by $T^* \approx 100$ K. The lower bound is 10 K for $V_g < 2.5$ V, where conductance approaches the noise floor at lower temperatures, and extends to 4 K for $V_g \geq 2.5$ V, where the conductance remains measurable above the noise floor.

The large uncertainty in $T_0$ at $V_g$ = 1V arises from increased conductance scatter within the 10 –100 K fitting window near threshold, where the exponentially suppressed conductance approaches the noise floor, reducing the precision of the VRH fit. The monotonic decrease of $T_0$ with gate voltage across the full range, despite the large uncertainty at the lowest gate voltage, is physically consistent with progressive filling of localized band-tail states and confirms the overall trend.

### S7.2 VRH Self-Consistency Check

A necessary condition for the physical validity of 2D Mott VRH is that the optimal hopping distance $R_{hop}$ exceeds the localization length ξ. For 2D Mott VRH, the optimal hopping distance [11,12] is:

$$R_{hop} = \frac{\xi}{3}(\frac{T_0}{T})^{\frac{1}{3}} \qquad \text{(S7.1)}$$

The condition R_hop > ξ requires $(\frac{T_0}{T})^{\frac{1}{3}} > 3$ , which is equivalent to the criterion T < $T_0$/27. When this criterion is violated, the VRH description breaks down; the extracted ξ then represents an upper bound.

For $V_g$ ≥ 3.3 V, $T_0$/27 falls below the lowest accessible measurement temperature (T = 4 K), meaning the self-consistency criterion $T_0$/27 > T is violated at every measured data point. The corresponding ξ values are therefore upper bounds, as shown in Table S2. For all other gate voltages ($V_g$ ≤ 3.2 V), $T_0$/27 ≥ 4 K, and the criterion is satisfied at the lowest measurement temperature; ξ values in this range are physically meaningful. The intrinsic localization length ξ = (3.5 ± 0.1) nm is extracted as the weighted mean over $V_g$= 1 –2.1 V, where $T_0$/27 ≫ 4 K.

Table S2. Summary of VRH parameters and ξ interpretation at selected gate voltages.

| $V_g$ (V) | $T_0$ (K) | ξ (nm) | ξ Interpretation |
|---|---|---|---|
| 1 | 11095 | 3 | Intrinsic |
| 2 | 5847 | 4 | Intrinsic |
| 3 | 250 | 18 | Intrinsic |
| 3.2 | 132 | 24 | Intrinsic |
| 3.3 | 99 | 28 | Upper bound |
| 4 | 10 | 88 | Upper bound |

**S8. Quantum capacitance correction to DOS $g(E_F)$**

The density of states formula used in the main text (Eq. 2) is the simplified expression valid when the gate dielectric capacitance $C_{hBN}$ dominates over the quantum capacitance $C_Q$ of the two-dimensional electron gas. Here we derive the full expression.

When a gate voltage $V_g$ is applied, it is distributed across two capacitances in series: the geometric capacitance $C_{hBN}$ of the hBN dielectric, and the quantum capacitance $C_Q$ of the 2D electron gas.

The total gate voltage partitions as:

$$V_g = V_{hBN} + V_Q \quad \text{(S8.1)}$$

where $V_{hBN} = {}^{Q}/_{C_{hBN}}$ is the voltage across the hBN dielectric and $V_Q = {}^{\Delta E_F}/_{e}$ is the voltage equivalent of the Fermi energy shift, with Q being the induced charge per unit area.

Since the two capacitors are in series, both carry the same induced charge Q. The quantum capacitance is defined as:

$C_Q = e^2 g(E_F)$ (S8.2)

where g(E_F) is the density of states at the Fermi level.

Finally, we get, the DOS, $g(E_F) = \frac{C_{hBN}}{e^2}[e / \left|\frac{dE_a}{dV_g}\right| - 1]$ (S8.3)

When $C_Q \ll C_{hBN}$, then $g(E_F) = \frac{C_{hBN}}{e} \left|\frac{dE_a}{dV_g}\right|^{-1}$ (S8.4)

In the plateau region (1.2–1.8 V), the measured derivative is $\left|\frac{dE_a}{dV_g}\right| \approx$ 8 meV/V = 0.008 eV/V. The inverse is $\left|\frac{dE_a}{dV_g}\right|^{-1}$ ≈ 125 V/eV. The correction term (−1 in absolute units of V/eV) is therefore negligible.

So, $\Delta g / g = \left|\frac{dE_a}{dV_g}\right|$ = 0.008 = 0.8% (S8.5)

### S9. Comparison of Localization Lengths in 2D Semiconductors

Table S3. Comparison of localization lengths across 2D semiconductors extracted from 2D variable-range hopping analysis.

| Sl No. | Material | Material layers/ thickness | Dielectric | Trap state density / DOS ($eV^{-1}$ $cm^{-2}$) | Temperature range (K) | ξ (nm) | Reference |
|---|---|---|---|---|---|---|---|
| 1 | $MoS_2$ | Monolayer to trilayer | $SiO_2$ | $4 \times 10^{12}$ | 30-295 | 8 | [13] |
| 2 | $MoS_2$ | 10 nm | $SiO_2$ | $(3.75\text{-}6.25) \times 10^{13}$ (DOS) | 6-80 | 3.4 (at -70 V) | [14] |
| 3 | $MoS_2$ | Monolayer | hBN | $10^{14}$ | 90-225 | 5 | [15] |
| 4 | $WS_2$ | Monolayer | $SiO_2$ | $2 \times 10^{14}$ (DOS) | 20-250 | 2 | [16] |
| 5 | $MoSe_2$ (MBE-grown) | Monolayer | Polymer electrolyte | $2 \times 10^{14}$ (DOS) | 80-200 | 9 | [17] |
| 6 | BP | 21.5 nm | hBN | $10^{12}$ | 10-230 | 11 | [18] |
| 7 | BP | 15 nm | $SiO_2$ | $10^{13}$ | 30-250 | 1 | [12] |

| 8 | $ReS_2$ | 10 nm | Polymer electrolyte | $4.17 \times 10^{14}$ (DOS) | 96-172 | 0.35-0.63 | [19] |
|---|---|---|---|---|---|---|---|
| 9 | $ReS_2$ | Few-layer | $SiO_2$ | $4 \times 10^{12}$ | 20–250 | 5 (at 15 V) | [20] |
| 10 | $ReS_2$ | 8 nm | hBN | $2.08 \times 10^{14}$ (DOS) | 10-200 | 3.5 | This work |

Trap state density values marked (DOS) indicate extraction via the density of states of the material, whereas unmarked values indicate extraction from the charge trap density at the dielectric interface.